\documentclass[doublecol]{epl2}
\usepackage{mathptmx,amssymb}
\usepackage{amsmath,ulem}
\def\CHG#1{{\textcolor{black}{#1}}}        % CHANGE

\usepackage{graphicx}
\usepackage{color}
\usepackage{epstopdf}
\usepackage{bm}

\title{Growth rate distribution and intermittency  in kinematic turbulent dynamos : which moment predicts the dynamo onset?}
\shorttitle{Growth rate distribution and intermittency  in kinematic turbulent dynamos } %Insert here a short version of the title if it exceeds 70 characters

\author{K. Seshasayanan, F. P\'etr\'elis}
\shortauthor{K. Seshasayanan, F. P\'etr\'elis}

\institute{                    
  Laboratoire de Physique Statistique, \'Ecole Normale Sup\'erieure,
PSL Research University; Universit\'e Paris Diderot Sorbonne Paris-Cit\'e; Sorbonne
Universit\'es UPMC Univ Paris 06; CNRS; 24 rue Lhomond, 75005 Paris, France.
}

\abstract{
We consider the generation of magnetic field by a turbulent flow. For the linear induction equation ({\it i.e.} the kinematic dynamo problem), we show that the statistical moments of the magnetic field display multiscaling and in particular moments of different order  turn unstable for different values of the control parameter. On a canonical example, we map the problem onto the calculation of the injected power by \CHG{a time correlated fluctuating force} acting on a Brownian particle. We are then able to calculate analytically the growth rate of the moments of the magnetic field and explain the origin of this intermittency. We finally show that the onset for the nonlinear problem is predicted by the linear onset of the moment of order $0^+$ (i.e. the logarithm of the magnetic field). }

\begin{document}

\maketitle

The dynamo effect is an instability that converts kinetic energy of an electrically conducting fluid into magnetic energy. It is the source of the magnetic field observed in most astrophysical objects, Earth and most planets, Sun and other stars, galaxies... This effect was identified by Larmor hundred years ago and yet many questions are still unanswered. In particular concerning the effect of turbulent fluctuations on the dynamo process. One of the first approaches to tackle this problem is the one of Kazantsev \cite{kaz} who modeled a turbulent flow as a delta-correlated in time process. A similar approach was made independently by Kraichnan to describe the evolution of a passive scalar \cite{krai}. Kazantsev studied the evolution of the magnetic energy, {\it i.e.} the second moment of the magnetic field. In that framework, several predictions were made depending on the spatial or temporal properties of the turbulent flow \cite{vincenzi, scheko1, aj}.  These models consist of a linear stochastic partial differential equation in which the stochastic term (that models the turbulent fluctuations) acts multiplicatively.  In a different context, the study of amplitude equations subject to noise  finds that multiplicative noises can create very intermittent behaviors which affect the moments of the field \cite{strata, onoff}. As a consequence, different moments grow with different growth rates. In such a case, the onset of  which moment predicts the dynamo threshold? 

To answer this question, we start with a numerical simulation of a turbulent dynamo in the class of Kazantsev dynamo. There have been very few numerical investigations of Kazantsev like dynamos. Most numerical studies  considered the dynamo instability by a flow  due to a random forcing in the Navier-Stokes equations \cite{scheko, mason}.  A numerical solution for the dynamo instability by a delta-correlated gaussian distributed velocity field was done in \cite{kanna}.  We use the same code  which is a modified version of \cite{mininni}.  The considered velocity field is of the form 
 ${\bf u} = {\boldsymbol \nabla} \times \left(  {\psi {\bf e}_z} \right) + u_z {\bf e}_z$,
%\begin{align}
%{\psi} & = U \zeta_1 \left( t \right) \Big( \sin \left( \phi_1 \left( t \right) %\right) \cos \left( k_f x + \phi_2 \left( t \right) \right) +  \cos \left( \phi_1 \left( t \right) \right) \sin \left( k_f y + \phi_2 \left( t \right) \right) \Big)/k_f, \label{Kazanflow1} \\
%u_z & = U \zeta_2 \left( t \right) \Big( \sin \left( \phi_1 \left( t \right) \right) \sin \left( k_f x + \phi_2 \left( t \right) \right) +  \cos \left( \phi_1 \left( t \right) \right) \cos \left( k_f y + \phi_2 \left( t \right) \right) \Big), \label{Kazanflow2}
%\end{align}
\begin{align}
{\psi} & = U \zeta_1 \left( t \right) \Big( \sin \left( \phi_1 \left( t \right) \right) \cos \left( k_f x + \phi_2 \left( t \right) \right)\nonumber\\
 & +  \cos \left( \phi_1 \left( t \right) \right) \sin \left( k_f y + \phi_2 \left( t \right) \right) \Big)/k_f, \label{Kazanflow1} \\
u_z & = U \zeta_2 \left( t \right) \Big( \sin \left( \phi_1 \left( t \right) \right) \sin \left( k_f x + \phi_2 \left( t \right) \right)\nonumber\\
&  +  \cos \left( \phi_1 \left( t \right) \right) \cos \left( k_f y + \phi_2 \left( t \right) \right) \Big), \label{Kazanflow2}
\end{align}
where $\zeta_1 \left( t \right), \zeta_2 \left( t \right)$ are two independent Gaussian white noise with  $\left\langle \zeta_1 \left( t \right) \zeta_1 \left( t' \right) \right\rangle_s = 2 D \delta \left( t - t' \right), \left\langle \zeta_2 \left( t \right) \zeta_2 \left( t' \right) \right\rangle_s = 2 D \delta \left( t - t' \right) $ with $\langle. \rangle_s$ the statistical average over realizations. $\phi_1 \left( t \right), \phi_2 \left( t \right)$ are two uniformly distributed random numbers in the interval $[0, 2 \pi]$.  We use the Stratanovich interpretation for the multiplicative terms that involve the noise \cite{strata}. 
This flow depends on two coordinates  so that  it is less computationally  expensive  to do statistics over long time series which allows us to obtain   accurate estimates of the higher order moments. 

We first consider the linear problem  (the induction equation) 
\begin{equation}
\partial_t {\bf B} = {\boldsymbol \nabla} \times \left( {\bf u} \times {\bf B} \right) + \eta \Delta {\bf B}\,,
\end{equation}
in which, using the independence of the flow on the $z$-direction, we write ${\bf B} = {\bf b} exp(i k_z z) + c.c.$. The governing equations are solved in a domain $\left[ 2\pi L, 2\pi L \right]$ with periodic boundary conditions. The field amplitude $B$ is defined as the square root of its energy $B^2=\overline{\bf B^2}$ where $\overline{f}$ stands for the spatial average. For $k_z L= 1$ and $k_f \, L = 4$, 
the magnetic energy is shown in fig. 1 and displays strong fluctuations. 

\begin{figure}
\begin{center}
\includegraphics[scale=0.17]{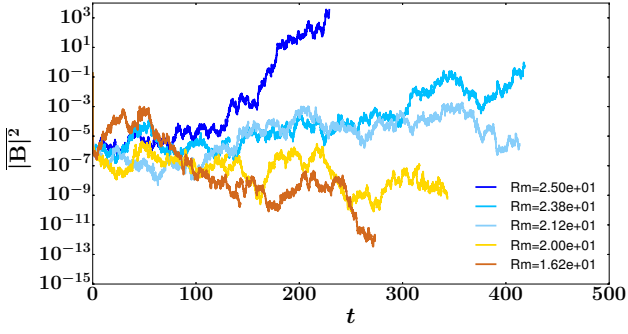}
\end{center}
\caption{The spatially averaged magnetic energy $B^2$ is shown as a function of time $t$ for different $Rm$ for the flow defined by eq. \ref{Kazanflow1}. }
\label{fig:timeseries1}
\end{figure}

We then calculate $\lambda_n$ the growth rate of the n-th moment of the magnetic field, defined by $\langle {B}^n \rangle_s\propto e^{\lambda_n t}$. 
$\lambda_n$ depends on  $Rm = U/\left( k_f \eta \right)$ and from a linear fit close to $\lambda_n=0$, we calculate the threshold of instability of each moment  denoted as $Rm_c (n)$.  We obtain   
$Rm_c(n=2) =  9.792  \pm 2.496$, $Rm_c(n=1) = 14.742 \pm 2.522$ and $Rm_c(n=0^+) =20.001 \pm 2.339$  (which corresponds to the log of the field, see discussion below)  \cite{errbar}. The onset of instability, calculated from the linear equation,  thus depends on the considered moment.

%%%%% On repart de la, ce paragraphe doit aller plus loin.

%We thus wonder if a similar phenomenon occurs for turbulent dynamos. In this letter, we consider a canonical example of flow in the class of Kazantsev dynamo. We are able to calculate the statistical properties of the growth rate of the moments of the field. We focus on their intermittent behavior and discuss their relevance for the realistic dynamo problem that takes into account nonlinear effects.  
 
 The flow in these numerical simulations involves many spatial scales and it is  thus difficult to derive analytical predictions for all the moments. \CHG{This can be done by considering a slightly different flow configuration.  
The velocity field is assumed to be an Ornstein-Uhlenbeck process (OU) in time and of the form ${\bf v}= Y(t) {\bf u}$, where ${\bf u}$ is a prescribed function of space. %with dimension of velocity. 
The noise $Y \left(t \right)$ satisfies $\langle Y (t) \, Y (t')\rangle_s= \exp(-|t-t'|/\tau) \, D/\tau  $. Here $\tau$ denotes the correlation time for the OU process. The white noise case is recovered in the limit $\tau \rightarrow 0$.  }%, where $D$ has the dimension of time and 
%$\langle.\rangle_s$ is the average over the realizations of the noise.  

Our strategy is %to consider that the flow is spatially periodic, and 
to use scale separation to obtain an equation for the part of the magnetic field that evolves at large scale compared to the scale of the flow.  
Noting again $\overline{f}$ the average over a wavelength of the flow, we write %$\langle.\rangle$ the average over a wavelength of the flow and write 
${\bf B} = \overline{\bf B}+ {\bf b}$. The fields satisfy  
\begin{eqnarray*}
\frac{\partial \overline {\bf B}}{\partial t}&=\nabla \times \left(\overline{ {\bf v} \times {\bf b}} \right) +\eta \nabla^2 \overline{\bf B} \,,
\label{EqbaseB}\nonumber\\
\frac{\partial {\bf b}}{\partial t}-\eta \nabla^2 {\bf b}& =\nabla \times \left( {\bf v} \times {\bf B} \right)-\nabla \times \left(\overline{ {\bf v} \times {\bf b}} \right) \,.
\end{eqnarray*}
In the framework of scale separation and  for ${\bf b}$ small compared to ${\bf B}$, the second equation writes
\begin{equation}
\frac{\partial {\bf b}}{\partial t}-\eta \nabla^2 {\bf b}=\nabla \times \left( {\bf v} \times \overline{\bf B}  \right)=\CHG{Y(t)}  \left( \overline{\bf B} \cdot \nabla {\bf u} - {\bf u} \cdot \nabla \overline{\bf B}  \right)\,.
%\label{Eqbsebsimple} 
\end{equation}

%The first term on the right corresponds to the $\alpha$-effect while the second term corresponds to the $\beta$-effect. 
\noindent To ease notation, we assume that the fields are $2 \pi$ periodic in all directions and  note $\hat{f}=(2\pi)^{-3/2}\int f e^{i {\bf k} r} d^3r$ the Fourier transform of $f$. % Its inverse reads $f=(2\pi)^{-3/2} \Sigma_{\bf k} \hat{f} \exp^{-i {\bf k} r}$, where the sum is performed over ${\bf k}=2\pi (n_x,\,n_y,\,n_z)$.
This  leads to 
\begin{equation}
\frac{\partial {\bf \hat{b}}}{\partial t}+\eta k^2 {\bf \hat{b}}= i  \CHG{Y(t)}  \Big(  \overline{\bf B}  \cdot \bf k \, {\bf \hat{u}} -  \hat{\bf u}  \cdot {\bf K} \overline{\bf B} \Big) \,,
\label{Eqbsebsimpleb} 
\end{equation}
with $k$ the norm of ${\bf k}$ and where $\overline {\bf B} $, the large scale field, is anticipated to be of the form  $\overline{\bf B}  \propto \exp{(i {\bf K . r })}$. We obtain the solution for ${\bf b}$ as 
\begin{equation}
{\bf \hat{b}} =   \CHG{W_k(t)}  \eta k^2 {\bf \hat{b}_r}\,,
\label{solb}
\end{equation}
where $\hat{\bf b}_r$ is the steady solution of Eq. \ref{Eqbsebsimpleb} with $\CHG{Y(t)}=1$, {\it i.e.} ${\bf \hat{b}_r}= i \Big(  \overline{\bf B} \cdot {\bf k} \,{\bf \hat{u}} -  \hat{{\bf u}} \cdot {\bf K} \overline{\bf B}  \Big)/(\eta k^2)$ and $\CHG{W_k(t)}$ is solution of 
\begin{equation}
\CHG{ \frac{dW_k}{d t}+\eta k^2 W_k= Y (t) \,.}
\label{EqYk}
\end{equation}
%$W_k$ is thus an Ornstein-Uhlenbeck process, with a damping rate $\eta k^2$. 
We then obtain the effect of the small scale fields on the large scale one as 
\begin{align}
\overline{{\bf v} \times {\bf b}}= & i \CHG{Y(t)} (2 \pi)^{-3} \sum_{\bf k} \CHG{W_k(t)} \Big(\,\overline{\bf B}  \cdot {\bf k}\; {\bf \hat{u}}(-{\bf k}) \times  {\bf \hat{u}}({\bf k})  - \Big. \nonumber \\
& \Big.  \, \hat{\bf u} \left( {\bf k} \right) \cdot {\bf K}\; \hat {\bf u} \left( - {\bf k} \right) \times  \overline{\bf B}  \Big) \,.
\label{formalpha}
\end{align}
If the velocity field contains modes with  wave vectors of same norm $k$, the expression is simplified and we obtain
\begin{eqnarray}
\overline{{\bf v} \times {\bf b}}&=&\CHG{Y(t) W_k(t)} \Big( \eta k^2 {\bm{\tilde{\alpha}}} \overline{\bf B}  \Big)\nonumber\\ &=&\CHG{Y(t) W_k(t)} \Big( \eta k^2 {\bm{\alpha}} \overline{\bf B} - \eta k^ 2 \bm{\beta} i {\bf K} \overline{\bf B}  \Big)
\label{formalphab}
\end{eqnarray}
where the tensor ${\bm{\tilde{\alpha}}}$ is obtained from   ${\bm{\alpha}}$ and $\bm{\beta} $ which are the alpha and beta tensor \cite{moff} that would be obtained with $Y(t)=1$, namely
\begin{align}
& \alpha_{pq}= (2 \pi)^{-3} i  \sum_{{\bf k},|{\bf k}|=k} \frac{{k}_q}{\eta {\bf k}^2} \left(\hat{{\bf u}}(-{\bf k})\times \hat{{\bf u}}({\bf k})\right)_p, \label{eqal1}\\
& \beta_{pqr} = ({2 \pi})^{-3}    \epsilon_{pmq} \sum_{{\bf k},|{\bf k}|=k} \hat{u}_m \left(- {\bf k} \right) \hat{ u}_r \left( {\bf k} \right). 
\label{formalphaclassique}
\end{align}
The formula \ref{formalphab}, \ref{eqal1} and \ref{formalphaclassique} give the expression of the $\alpha$-tensor for  the  random flow that we consider.  We note that it has properties less simple than assuming $\alpha$ to be a gaussian white noise, a standard way to include fluctuations in a mean field model, see for instance \cite{branden}. In particular, we will find that the distribution of the fluctuations are non gaussian.

The antisymmetric part of the $\bm{\tilde{\alpha}}$-tensor leads to an advection of the field and does not affect the growth rate. We thus consider a symmetric tensor. 
We then change coordinates to diagonalize it so that ${\bm{\tilde{\alpha}}} \overline{\bf B} =(\alpha_1 \overline{\bf B}_1,\alpha_2 \overline{\bf B}_2, \alpha_3 \overline{\bf B}_3) $. The most unstable mode is obtained by finding, among the $\alpha_i$ of same sign, the two largest $|\alpha_i|$, say $\alpha_1$ and $\alpha_2$ and considering a magnetic field of the form $\overline{\bf B}=\hat{\bf B} e^{-i K z}$. 
%Considering a magnetic field of the form $\langle {\bf B} \rangle=\bar{B} e^{-i K z}$, we change coordinate to diagonalize it so that $\bm{\alpha} \langle {\bf B} \rangle =(\alpha_1 \langle {\bf B} \rangle_1,\alpha_2 \langle {\bf B} \rangle_2, 0 ) $. The $\bm{\beta}$-tensor acts like a diffusion term hence it is already diagonal. %The most unstable mode is obtained by finding the two largest $|\alpha_i|$, say $\alpha_1$ and $\alpha_2$.
For positive $\alpha_{1,2}$, the eigenmode $B_p=\sqrt{\alpha_1} \hat{B}_1+i \sqrt{\alpha_2} \hat{B}_2$ satisfies 
\begin{equation}
\frac{d B_p }{d t}= \CHG{Y(t) W_k(t)} \alpha \eta k^2 K B_p-\eta K^2 B_p \,.
%\frac{d B_p }{d t}= \CHG{Y(t) W_k(t)} \eta k^2 \Big( \alpha K ) B_p-\eta K^2 B_p \,.
\label{EqBp}
\end{equation}
where $\alpha=\sqrt{\alpha_1 \alpha_2}$. %, $\beta = \left( \hat{e}_z \cdot {\bf u} \right)^ 2/\left( \eta k^2 \right)$ 
 We then obtain the large-scale magnetic field as
\begin{equation}
B_p(t)=B_p(0) e^{\alpha \eta k^2   K   I(t) -\eta K^2 t} \,.
\label{EqBpint}
\end{equation}
where  $I(t)=\int_0^t \CHG{W_k(t') Y(t')} dt'$.

The magnetic field has thus a fluctuating growth rate controlled by the random variable $Y(t)$.  It is then  pleasant that this quantity was studied  by Jean Farago \cite{farago2}. Indeed, the velocity  of a Brownian particle subject to a random force in a viscous fluid satisfies Eq. \ref{EqYk} where $\CHG{Y(t)}$ is the random force and $\eta k^2$ the viscous damping rate. The quantity of our desire, $I(t)$, is then the energy injected by the random force into the particle.  This quantity follows a law of large deviation and at long time its probability density function  takes the form 
\begin{equation}
P(I=t \epsilon)\simeq e^{-t g(\epsilon)}\,.
\label{defrate}
\end{equation}
where $\simeq$ means that the log are equivalent and $g$ the  rate function is given for positive energy by \CHG{its Legendre transform $h$ as \cite{farago2} 
\begin{align}
g \left( \epsilon \right) = & h \left( \gamma \right) - \gamma \epsilon\,, \label{resgY0} \\
h' \left( \gamma \right) = & \epsilon, \label{gammadef}
\end{align}
}and is infinite for negative $\epsilon$ \cite{Rem}. \CHG{ The Legendre transform is given by 
\begin{align}
h \left( \gamma \right) = & \frac{1}{2 \tau} \Bigg( - \eta k^2 \tau - 1 + \Big( \eta^2 k^4 \tau^2 + 1 + 2 \eta k^2 \tau \Big. \nonumber \\
& \Big.  \sqrt{ 1 + \frac{4 D \gamma}{\eta k^2} } \;\;\; \Big)^{\frac{1}{2}} \Bigg).
\end{align}
Then $\gamma$ is found by inverting $h' \left(\gamma \right) = \epsilon$. }

We can now calculate the growth rate of the moments of the magnetic field. As the small scale field ${\bf b}$ is small compared to the large scale one, the spatially averaged magnetic energy is proportional to  $\overline{\bf B}^2$ and we have   %  such that $\langle {\bf B}^n \rangle_s\propto e^{\lambda_n t}$. We have
\begin{align}
\langle {B}^n \rangle_s \propto & \int e^{ n \eta k^2  \alpha K  \epsilon t-\eta n K^2 t} P(\epsilon) d\epsilon \simeq \nonumber \\ 
&\,\int e^{-t \left(g(\epsilon)- n \eta k^2  \alpha K  \epsilon\right) -\eta n K^2 t} d\epsilon.
\label{Eqmoment}
\end{align}

For large $t$, this is  evaluated by Laplace method. \CHG{ Let $\epsilon_c (n), \gamma_c(n)$ be solution of $g'(\epsilon)= - \gamma = n \alpha \eta k^2 K$, the growth rate of the n-th moment is 
\begin{align}
\lambda_n= & - g\left( \epsilon_c \right) + n \epsilon_c \eta k^2 \alpha K -n \eta K^2, \nonumber \\
= & - h\left( \gamma_c \right) - n \eta K^2\,,
\end{align} 
where we have used equation \ref{resgY0} to replace $g\left( \epsilon \right)$ by $h\left( \gamma \right)$. }

%A positive growth rate requires a negative $K\alpha_1 \alpha_2$ which is assumed from now on.

%We find that $i_c$ is bounded if $4 n D K \alpha \le 1$.

%{\bf That is weird! If this is not verified, because $g$ decreases only linearly at large $i$, there is a value of $n \alpha$ above which the integrals in Eq. \ref{Eqmoment} diverge. What does it mean? I tried to inject next order in the large deviation expansion but it did not help.}

\noindent \CHG{Provided $\gamma_c$ is real, we obtain
\begin{align}
\lambda_n=-n \eta K^2+\frac{1}{2 \tau} \Bigg( & 1 + \eta k^2 \tau - \Big( 1 + \eta^2 k^4 \tau^2 \Big. \Bigg. \nonumber \\
\Big. \Bigg. & +2 \eta k^2 \tau \sqrt{1 - 4 D n \alpha K} \;\; \Big)^{\frac{1}{2}} \Bigg)\,. \label{eqmoments}
\end{align}
We also find that the logarithm of $B$  grows like $ \left( \eta k^2 \alpha K   D/(1+\eta k^2 \tau)-\eta K^2 \right) t$.}\\ 
 We note that the behavior of the log can be obtained directly from the behavior of the moments (here Eq. \ref{eqmoments}). Indeed, for $n \rightarrow 0$, a standard heuristic estimate of statistical mechanics  writes  $\langle {B}^n \rangle_s\simeq 1+ n \langle \log({B}) \rangle_s$ and $e^{\lambda_n t}\simeq 1+\lambda_n t$, so that $\langle \log({ B}) \rangle_s/t$ tends to $\lim_{n \rightarrow 0} \lambda_n/n$. This can be checked for Eq. \ref{eqmoments}. We thus say that the increase or decrease  of the log  of the field is obtained from the sign the growth rate of the moment of order $0^+$.

%The onset defined with the one of the logarithm of $B_p$ is \CHG{as} $k^2 \alpha D/|K|=1$.

The moments display multiscaling: their growth rates vary non linearly with the order $n$. \CHG{In the limit of infinite $Rm$, this has been predicted  for  random renovating flows  \cite{molch, almighty} and for linear flows \cite{chertkov}.} Similar predictions were also made but restricted to a few first  moments of integer order: in the case of a linear shear combined with a random nonhelically forced flow, it was shown analytically that the first and second moments have different growth rates \cite{scheko}. This was also shown for the first four moments of  a mean field model, using a random alpha-effect \cite{branden}.
 
From Eq. \ref{eqmoments}, we observe that  the onset defined by the vanishing value of the  growth rate depends on $n$. More precisely, the onset of the n-th moment behaves as $k^2 \alpha_c (n) D/|K|\simeq 1 + \eta k^2 \tau - n (K/k)^2 (1+3\eta k^2 \tau + \eta^2 k^4 \tau)$. Moments for large $n$, larger than $(4 D K \alpha)^{-1}$, diverge faster than exponentially. \CHG{The limit $\tau \rightarrow 0$ leads to the velocity field being uncorrelated in time (white noise limit), which ressembles \CHG{the Kraichnan-Kazantsev class of velocity fields}. Anticipating from the $0^+$-moment the value of the onset, we get at first order in $\tau$ the onset of the dynamo instability to be $\alpha_c (0) \simeq K ( 1 + \eta k^2 \tau)/(k^2 D)$. We find that  increasing $\tau$ the time correlation of the velocity field leads for small $\tau$ to a larger threshold for the dynamo instability. In addition the onset for the $n$-th moment in this limit is given up to first order in both $\tau, (K/k)^2$ by, $ k^2 \alpha_c (n) D/|K| \simeq 1 + \eta k^2 \tau - n (K/k)^2 (1+3\eta k^2 \tau)$. We see that the difference in the threshold for the growth of different moments $|\alpha_c(n) - \alpha_c(m)|$ increases with increasing $\tau$. } \CHG{To sum up, in the $\tau \rightarrow 0$ limit, the dynamo instability threshold and the multiscaling increase with increasing correlation time $\tau$. This result is non-trivial and it is important to note that the velocity field in the analytical study has a zero mean.}
%Nonetheless, it would be interesting to find other examples of flows where a highly fluctuating in time velocity field has a dynamo threshold that decreases with decreasing correlation time. }

% \CHG{An explanantion for the increase in threshold for increasing $\tau$ can be understood by looking at the amplification factor $\epsilon_c$ (the injected power in the statistical physics analogy). We find that in the $\tau \rightarrow 0$, $\epsilon_c \simeq D( 1/\sqrt{1 - 4 Dn K \alpha} - \eta k^2 \tau/2)$ decreases as one increase $\tau$. Thus one needs higher $\alpha$ to overcome joule dissipation. }
%(@ write the part below kana)

%There have been very few numerical investigations of Kraichan-Kazantsev like dynamos. Most numerical studies  considered the dynamo instability by a flow  due to a random forcing in the Navier-Stokes equations \cite{scheko, mason}.  A numerical solution for the dynamo instability by a delta-correlated gaussian distributed velocity field was done in \cite{kanna}.  We use the same code  which is a modified version of \cite{mininni}. %The Stratanovich calculus for the noise is used, (see \cite{greiner1988numerical,leprovost2004influence}). 
To test our analytical predictions, we consider \CHG{a delta-correlated in time flow} of the Roberts type \cite{roberts} defined as 
\begin{equation}
{\bf v} = \zeta(t) U \left( cos (k y), sin (k x), cos ( k x) + sin(k y) \right)\,.
\label{eqnoiseroberts}
\end{equation}
 %The governing equations are solved in a domain $\left[ 2\pi L, 2\pi L \right]$ with periodic boundary conditions. We choose  a flow depending on $2$ coordinates  as it is less computationally  expensive thus allowing us to do statistics over a long time series. This eventually helps us to have more accurate estimates of the higher order moments. 

%We show in fig.  \ref{fig:Growth1} the total magnetic energy defined as $\left\langle |{\bf B}^2| \right\rangle = \int\int {\bf B}^{\dagger} {\bf B} \,\, dx dy$ as a function of time $t$ with

\begin{figure}
\begin{center}
\includegraphics[scale=0.15]{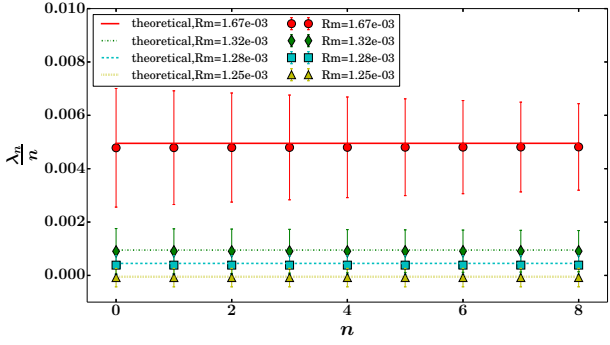}
\end{center}
\caption{For the flow defined by Eq. \ref{eqnoiseroberts}, growth rate $\lambda_n/n$ is shown as a function of $n$ for different values of $Rm$. The parameter $K/k = 0.0025$.}
\label{fig:Growth1}
\end{figure}

\noindent For $K/k = 0.0025$, % for different values of $Rm$ as mentioned in the legend. 
we calculate the growth rate $\lambda_n$ of the moments of the magnetic field from the numerical solution \cite{treat}. Figure \ref{fig:Growth1} shows $\lambda_n/n$ as a function of $n$ for different values of $Rm$ defined as $Rm = U/(\eta k)$. %The $n=0^+$ value ($\lim_{n \rightarrow 0} \lambda_n/n$) gives the growth of logarithm of the magnetic field. %The analytical solutions from equation \ref{eqmoments} are shown in solid curves, the numerical results are shown by the data points. 
The numerical results and the theoretical solutions  agree very well. We note that $\lambda_n/n$ stays constant for different values $n$ hence the growth rate of the  moments $\lambda_n$ scales linearly in $n$. %The nonlinear scaling is observed for very large $n \sim (K/k)^{-2}$.

\begin{figure}
\begin{center}
\includegraphics[scale=0.15]{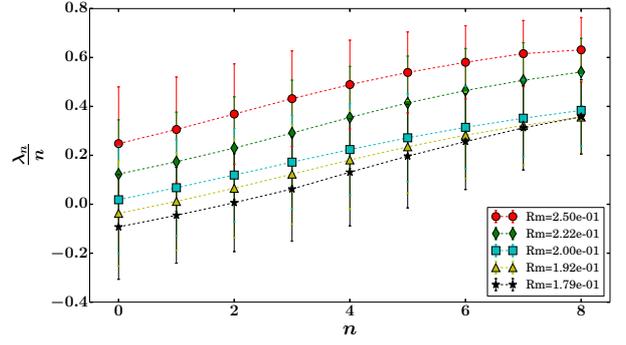}
\end{center}
\caption{For the flow defined by Eq. \ref{eqnoiseroberts}, growth rate $\lambda_n/n$ is shown as a function of $n$ for different values of $Rm$. The parameter $K/k = 0.25$.}
\label{fig:Growth2}
\end{figure}

In order to observe a nonlinear scaling near the threshold we need to reduce the scale separation. Indeed expanding Eq. \ref{eqmoments} for the flow studied here \CHG{(with $\tau = 0$)}, we obtain
\begin{align}
\lambda_n= & -n \eta K^2+ \eta k^2 \Big( n \frac{D U^2}{\eta} \frac{K}{k} (1 - \frac{K}{k}) + \nonumber \\
& n^2 \frac{D^2 U^4}{\eta^2} \frac{K^2}{k^2} (1 - \frac{K}{k})^2 + \cdots \Big)\,.
\label{linscal}
\end{align}
Anticipating again from the $0^+$-moment the value of the onset to be $DU^2/\eta = K/k (1-K/k)^{-1}$, we obtain at onset $\lambda_n = \eta k^2 n^2 K^4/k^4$. Thus for $K/k \ll 1$ and $n \sim O(1)$ the growth rate $\lambda_n$ scales linearly with $n$, but as we increase $K/k$ we  start to see contributions from higher orders of $n$. Using $K/k = 0.25$, we show $\lambda_n/n$ as a function of $n$ for different values of $Rm$ in figure \ref{fig:Growth2}. The scaling of $\lambda_n$ is nonlinear with respect to $n$. \CHG{The theoretical results, not displayed here, are not valid  as they assume large scale separation. There have been many studies which have considered the validity of the first order smooting approach in the context of alpha-effect \cite{brandy, Rogy}. It has recently been studied in detail by \cite{shum}, where it is  shown that for small scale separation there is a significant difference from the theoretical growth rate. } 

The results presented so far show that the moments calculated from the linear induction equation have different onsets.  The reader familiar with usual instabilities should be worried at that stage. This paradoxical behavior is actually reminiscent of  bifurcating systems in the presence of multiplicative noise. Consider the canonical model $\dot{x}=(\mu+\zeta(t)) x-x^3$ where $\mu$ is the control parameter and $\zeta$ a white noise of autocorrelation $D \delta(t)$. Dropping the nonlinear term, the solution reads $x(t)=x(0)\exp^{(\mu t +\int_0^t \zeta(t') dt')}$ so that $\langle x^n \rangle_s \propto \exp^{(n \mu t+n^2 D t/2)}$. The onset of the n-th moment is given by $\mu_c=-n D/2$. This traces back to the intermittent  behavior of $x$: there exists, on rare occasion, coherent occurences of the noise during which $x$  keeps on growing exponentially for long durations. These phases provide large contributions for large moments of the field and are responsible for the decrease of $\mu_c$ as a function of $n$ \cite{onoff}. It is important to realize that these events are suppressed as soon as a nonlinearity is taken into account. Indeed, the Fokker-Planck equation for the nonlinear model can be solved analytically. It shows that for negative $\mu$, $x$ tends to $0$ and that this solution is unstable for positive $\mu$. The onset when taking nonlinearities into account is thus $\mu_c=0$. It is given by the onset of the n-th moment of the linear problem when $n$ tends to zero. In other words, the onset corresponds to the onset of the logarithm of the field, the Lyapunov, when calculated from the linear equation. This result holds even for extended systems \cite{Ptheo}. In the context of  kinematic turbulent dynamo, the onset is thus given by the change of sign of the variation of the statistical average of the log of the magnetic field, and not by the behavior of any other statistical  moment, in particular, not by the one of $n=2$ associated to the energy of the field.

%%%%%%%%%%%
%Repartir de la

\begin{figure}
\begin{center}
\includegraphics[scale=0.187]{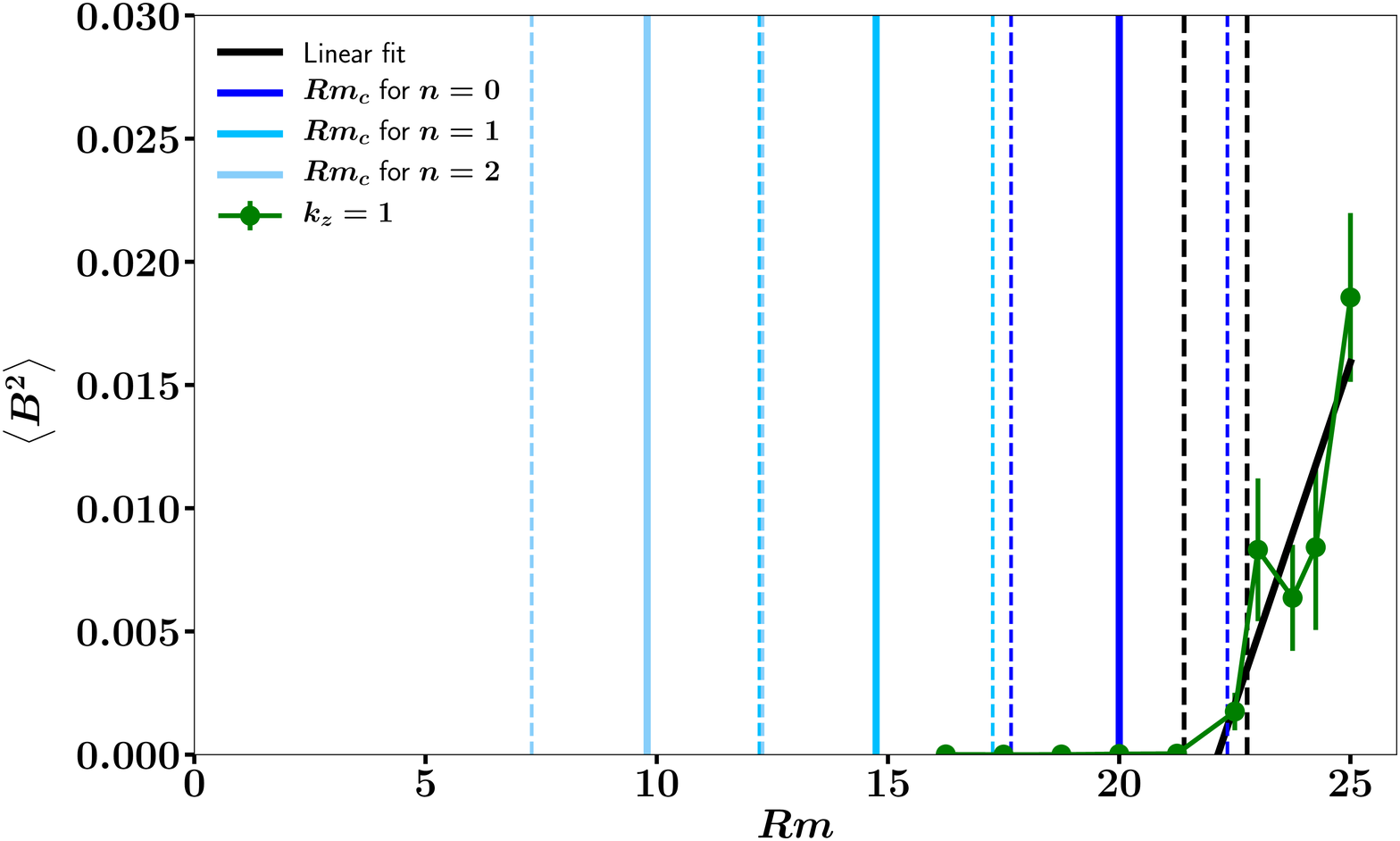}
\end{center}
\caption{The time averaged magnetic energy $\left\langle B^2 \right\rangle$ from the solutions of Eq. \ref{eqjb} is shown for different values of $Rm$. The continuous vertical lines are  the predictions of threshold $Rm_c$ for different moments $n$ using the kinematic simulations. The dotted lines display the error bar. }
\label{fig:Satnew}
\end{figure}

To test this prediction we have performed numerical simulations that include nonlinear effects for the magnetic field. If we solve for the full magnetohydrodynamic  system of equations we need to solve a $3D$ flow as the nonlinearity makes the $2D$ problem become $3D$. In order to remain computationally efficient we have  considered several simpler forms of nonlinearity. For the flow defined by eq. \ref{Kazanflow1}, we have solved
\begin{align}
\partial_t {\bf B} = {\boldsymbol \nabla} \times \left( {\bf u} \times {\bf B} - \left\langle |{\bf B}|^2 \right\rangle_z {\bf J} \right) + \eta \Delta {\bf B},
\label{eqjb}
\end{align}
where ${\bf J} =\frac{1}{\mu_0} \left( {\boldsymbol \nabla} \times {\bf B} \right)$ is the current. %We calculate the total magnetic energy $\left\langle |{\bf B}|^2 \right\rangle$ for different values of $Rm = U/ \left( k_f \eta \right)$. 
%We do a linear fit of the magnetic energy as a function of $Rm$ to find the threshold. This is the correct threshold as given by the nonlinear equation and is compared with the kinematic dynamo results $Rm_c \left( n \right)$.  
We show the amplitude of the space and time averaged magnetic energy $\left \langle   B^2 \right\rangle$ as a function of $Rm= U/ \left( k_f \eta \right)$ in figure \ref{fig:Satnew}.  The solid dark line denotes a linear fit through the data points. The x-intercept of the linear fit is the actual threshold of the dynamo instability.  We have $Rm_c(NL)=22.082   \pm 0.623$. The error bars of the x-intercept of the linear fit which gives the error in calculating the threshold of the instability  is found using  a bootstrapping algorithm. 
%The threshold from the kinematic dynamo results for different moments $n$ are shown by vertical lines. 
%The vertical and horizontal error bars  are calculated using standard bootstrap algorithm with $95\%$-confidence interval \cite{bootstrap}. We obtain  {\bf (mettre les bonnes valeurs)} 
%$Rm_c(n=0) = 0.1967 \pm 0.0016$, 
%$Rm_c(n=1) = 0.1895 \pm 0.0024$, 
%$Rm_c(n=2) = 0.1717 \pm 0.0035$. .  
 Compared with $Rm_c(n)$ as discussed initially,  we conclude that the value for the $0^+$-moment is equal within error bar to  $Rm_c(NL)$  while the energy ($2$-order moment) underestimates  the threshold.  

It is important to realize that the form of the nonlinear term does not change the value of the onset but that without nonlinear term, different moments have different onsets. We have checked this by considering two other nonlinear terms. For the flow defined by eq. \ref{eqnoiseroberts}, we have solved 
\begin{equation}
\frac{\partial {\bf B} }{\partial t}=\nabla \times \left( {\bf v} \times {\bf B} - \left\langle |{\bf B}|^2 \right\rangle_z {\bf B} \right) +\eta \nabla^2  {\bf B} \,.
\label{Nonlin1}
\end{equation}
Using the same data analysis as for the former flow we obtain 
%The time series of the magnetic field is shown in figure \ref{fig:Sattime1}  for different values of $Rm$. We can see that the field is highly intermittent. We show the amplitude of the time averaged magnetic field $\left\langle \overline{|\bf B|^2} \right\rangle$ as a function of $Rm$ in figure \ref{fig:Sat1}.  The solid dark line denotes a linear fit through the data points. The x-intercept of the linear fit is the actual threshold of the dynamo instability. 
%The threshold from the kinematic dynamo results for different moments $n$ are shown by vertical lines. 
%The vertical and horizontal error bars  are calculated using standard bootstrap algorithm with $95\%$-confidence interval \cite{bootstrap}. We obtain 
 for the kinematic simulation $Rm_c(n=0^+) = 0.1967 \pm 0.0016$, 
$Rm_c(n=1) = 0.1895 \pm 0.0024$, 
$Rm_c(n=2) = 0.1717 \pm 0.0035$ while with the nonlinear term $Rm_c(NL)= 0.1970 \pm 0.0011 $.  
 %As we can see the prediction from the $0$-moment captures the instability clearly while the $2$-order moment underestimates  the threshold. 

%\begin{figure}
%\begin{center}
%\includegraphics[scale=0.15]{figs_04.eps}
%\end{center}
%\caption{The amplitude of the magnetic energy $\left\langle |{\bf B}^2| \right\rangle$ as a function of time $t$ for the solutions of Eq. \ref{Nonlin1}.}
%\label{fig:Sattime1}
%\end{figure}
%\begin{figure}
%\begin{center}
%\includegraphics[scale=0.15]{figs_05.eps}
%\end{center}
%\caption{The time averaged magnetic energy $\left\langle \overline{\bf |B|^2} \right\rangle$ from the solutions of Eq. \ref{Nonlin1}
%is shown for different values of $Rm$. The continuous vertical lines are  the predictions of threshold $Rm_c$ for different moments $n$ using the kinematic simulations. The dotted lines display the error bar. }
%\label{fig:Sat1}
%\end{figure}

We finally solved for  a last form of nonlinear term,  introduced by considering the full set of Navier Stokes along with the induction equation.   
The velocity field is forced by a forcing which is random in time of the form ${\bf f} = \zeta(t) f_0 \left( cos (k y), sin (k x), cos ( k x) + sin(k y) \right)$ with $\langle \zeta(t)\zeta(0)\rangle_s= \delta(t)$. The governing equations are
\begin{align}
\partial_t {\bf v} + {\bf v} \cdot \nabla {\bf v} & = - \frac{1}{\rho} \nabla p + \nu \,\Delta {\bf v} + {\bf f} - \frac{1}{\rho} \left\langle \left( {\bf J} \times {\bf B} \right)  \right\rangle_z \label{eqn:fullsys1} \\
\partial_t {\bf B} & = \nabla \times \left( {\bf v} \times {\bf B} \right) + \eta \, \Delta {\bf B}. \label{eqn:fullsys2}
\end{align}
Here ${\bf J} = \frac{1}{\mu_0} \nabla \times {\bf B}$ is the current, $\left\langle \cdot \right\rangle_z$ denotes averaging along the $z$-direction. %\sout{The Lorentz force is averaged along the $z$-direction to maintain the independence of the flow on $z$. We point out that this set of equations is the limit of the Navier Stokes and induction equation in the limit of infinite rotation} \CHG{
These equations can be obtained from the Navier-Stokes and the induction equation in the limit of infinite rotation \cite{gallet}. Only the $z$-independent component of the Lorentz force is considered because  the $z$-dependent component induces a correction in the velocity field proportional to the inverse of the rotation rate and hence can be neglected.  
 We define $Rm = \sqrt{f_0/k}/(\eta k)$ to be the magnetic Reynolds number. For the parameters $K/k = 0.25,\, Re = \sqrt{f_0/k}/(\nu k) = 0.05$, using the same data analysis as for the former flow we obtain  
$Rm_c(n=0^+) = 4.0272 \pm 0.0183$,  
$Rm_c(n=1) = 3.9042 \pm 0.0501$, 
$Rm_c(n=2) = 3.7865 \pm 0.0515$, 
$Rm_c(NL)  = 4.0151 \pm 0.0102$. 
The same conclusions as for the other nonlinear models apply. In particular the threshold is given by the $0^+$-moment of the magnetic field and higher moments underestimate the threshold. We point out that  the velocity field is not a delta-correlated process and has a finite correlation time. 
Together with the analytical prediction of Eq. \ref{eqmoments}, these numerical results  show that  intermittency and multiscaling of the moments together with the existence of different onsets of instability for different moments, is not a property restricted to delta correlated velocity fields but is generic to any fluctuating flows \cite{specfreq}.

All the models  investigated here display strong intermittency with the growth rate of the n-th moment depending non linearly on $n$. In particular the threshold of instability calculated from the linear equation depends on $n$. When nonlinear effects are considered, the threshold becomes uniquely defined and is provided by the vanishing of the linear growth rate of the  log of the field (statistical moment of order $n=0^+$).  These properties are expected to hold for all turbulent dynamos. Numerical simulations of the linear induction equation do not frequently consider statistical averages but instead measure the evolution of the log of the magnetic energy ($B^2=\overline{\bf B^2}$) so that a long time decay (respectively growth) of the log amounts to a decay (resp. growth) of its statistical average (see \cite{treat} for a relation between times series of the kinematic problem and statistical averages). This thus correctly predicts the onset. If instead a statistical average of the magnetic energy (or any other moment different from the log) is made, then the predicted onset would be wrong. Similarly, we point out that most studies on Kazantsev dynamo focus on the $n=2$ moment and are likely to give at best an approximation of the onset.  

Out of the dynamo context, it seems worth investigating whether similar behaviors play a role in other systems with multiplicative noise, such as the advection of a passive scalar by a turbulent flow.

Finally, our results draw a link between a highly out of equilibrium system (turbulent dynamo) and a classical example of stochastic process (Brownian particle). This has two interesting consequences.  First other tools of statistical mechanics can be used to study the dynamo in that context, in particular instanton methods and  concentration of measure. Second, a similar approach is expected to be of interest in a variety of problems when scale separation can be used, including but not restricted to other hydrodynamic instability such as the anisotropic kinetic alpha effect (aka) for instance \cite{aka}. \\   
 %(other examples?)

%(@ i will write the following)
%T%he onset depends on the moment?
%What the meaning of that?
%Reminiscent of simple SDE.

%What for an extended system? The lyapunov

%We need nonlinearity.

%Nonlinear code (@write this part kana). model A and model B

%i%gure of the moments

%(@ i will write the following)
%Conclusion

%\begin{figure}[htb!]
%\begin{center}
%\includegraphics[width=80mm]{mecanism_b.eps}
%\caption{Sketch of the different steps involved in the amplification mechanism $\alpha^{\sigma}$ for a typical geophysical flow. Top: Two adjacent convective cells (grey %cylinders) with axial vorticity $\omega$ are subject to a transverse azimuthal magnetic field  $B$ (red). Middle: Both upward and downward axial currents $J\propto ({\bf v \times  B})$ (blue) are induced between the convective cells. Bottom: In presence of conductivity gradients correlated to the vorticity (maximum gradient  represented by pink dashed lines), large (resp. low) conductivity increases (resp. decreases) the induced current: the resulting net upward current $J^{'}$ is parallel to the vorticity. 
%}
%\label{fig2}
%\end{center}
%\end{figure}
F.P. thanks W.R. Young for several enlightening discussions on the large deviation function  of the injected power to a Brownian particles.


\begin{thebibliography}{0}
\bibitem{errbar} The error bars  are calculated using standard bootstrap algorithm with $95\%$-confidence interval \cite{bootstrap}.
\bibitem{Rem} Eq. \ref{resgY0} is obtained for a bounded initial condition for $W$. Unboundeed distributions of the initial condition change the rate function \cite{farago2, farago} but this concerns negative or small values of the injected power and is not relevant for the calculation of the moments $n \ge 0$.
\bibitem{treat} The average over realizations can be done by performing several independent simulations. Another possibility is to use a long time series of duration say $T$. By splitting it in $N$ shorter time series (duration $\tau=T/N$) and rescaling the amplitude to fix its initial value, we generate $N$ independent realizations provided that $\tau$ is larger than the correlation time of the fields.
\end{thebibliography}
\end{document}